\documentclass[twocolumn,preprintnumbers,showpacs,amsmath,amssymb]{revtex4}
\usepackage{graphicx}
\usepackage{amssymb}

\begin{document}

\title{Unconditionally secure quantum bit commitment based on the
uncertainty principle}
\author{Guang Ping He}
\email{hegp@mail.sysu.edu.cn}
\affiliation{School of Physics, Sun Yat-sen University, Guangzhou 510275, China}

\begin{abstract}
Unconditionally secure quantum bit commitment (QBC) was considered
impossible. But the no-go proofs are based on the Hughston-Jozsa-Wootters
(HJW) theorem (a.k.a. the Uhlmann theorem). Recently it was found that in
high-dimensional systems, there exist some states which can display a chaos
effect in quantum steering, so that the attack strategy based on the HJW
theorem has to require the capability of discriminating quantum states with
very subtle difference, to the extent that is not allowed by the uncertainty
principle. With the help of this finding, here we propose a simple QBC
protocol which manages to evade the no-go proofs.
%
\end{abstract}


\maketitle


\section{Introduction}

Besides the well-known quantum key distribution \cite{qi365}, bit commitment
(BC) is another essential cryptographic primitive. It is a two-party
cryptography including two phases. In the commit phase, Alice decides the
value of the bit $b$ ($b=0$ or $1$) that she wants to commit, and sends Bob
a piece of evidence, e.g., some quantum states. Later, in the unveil phase,
Alice announces the value of $b$, and Bob checks it with the evidence. An
unconditionally secure BC protocol needs to be both binding (i.e., Alice
cannot change the value of $b$ after the commit phase) and concealing (Bob
cannot know $b$ before the unveil phase).

BC is closely related with many other cryptographic tasks, e.g., coin
tossing \cite{qi365} and oblivious transfer\ \cite{qbc9,*qbc185}, all of
which are the building blocks for even more complicated multi-party secure
computation protocols \cite{qi139}. Unfortunately, it is widely accepted
that unconditionally secure quantum BC (QBC) is impossible \cite%
{qi74,qi24,qi23,qi56,qi58,qi105,qi82,qbc28,qi47,qi611,qi157,qi147,qbc49,qi101,qbc3,qbc36,qbc35,qi610,qi240,qi283,qi323,qi715,qbc12,qi714,qbc32,qbc31,qbc121,qbc140,qbc146,qbc148,qbc162}
, despite of some attempts towards secure ones (a detailed list and brief
history can be found in the introduction of \cite{HeJPA}). This result,
known as the Mayers-Lo-Chau (MLC) no-go theorem, was considered as putting a
serious drawback on quantum cryptography.

But all these no-go proofs are based on the Hughston-Jozsa-Wootters (HJW)
theorem \cite{qi73}, which also appeared under different names (e.g., the
Uhlmann theorem) and presentations in literature \cite%
{qbc48-2,qbc8,qbc48-3,qi1454,qi1455,qbc186,qbc48}. As concisely summarized
in \cite{qi56}, the conclusion of the HJW theorem can be expressed as
follows.

\bigskip

\textit{The HJW theorem:}

\textit{Let }$\psi _{1}^{0}$\textit{,\ }$\psi _{2}^{0}$\textit{,\ ..., }$%
\psi _{m}^{0}$\textit{\ and }$\psi _{1}^{1}$\textit{,\ }$\psi _{2}^{1}$%
\textit{,\ ..., }$\psi _{m^{\prime }}^{1}$\textit{\ be two sets of possible
quantum states with associated probabilities described by an identical
density matrix }$\rho $\textit{. It is possible to construct a composite
system }$\alpha \otimes \beta $\textit{\ such that }$\beta $\textit{\ alone
has density matrix }$\rho $\textit{\ and such that there exists a pair of
measurements }$M_{0}$\textit{, }$M_{1}$\textit{\ with the property that
applying }$M_{0}$\textit{\ (}$M_{1}$\textit{) to }$\alpha $\textit{\ yields
an index }$i$\textit{\ of state }$\psi _{i}^{0}$\textit{\ (}$\psi _{i}^{1}$%
\textit{) to which }$\beta $\textit{\ will have collapsed.}

\bigskip

However, it is worth noting that there is one thing left unanswered in the
HJW theorem: will the two measurements $M_{0}$ and $M_{1}$ also be
sufficiently distinguishable? Very recently, it was found \cite{HeSteering}
that if we take $\{\psi _{i}^{0}\}$\ and $\{\psi _{i}^{1}\}$ as the two sets
of evenly distributed states%
\begin{equation}
\left\{ \psi _{i}^{0}\equiv \left\vert \phi _{i+}\right\rangle =\frac{1}{%
\sqrt{2}}(\left\vert 0\right\rangle +\left\vert i\right\rangle
),i=1,...,n-1\right\}   \label{state+}
\end{equation}%
and%
\begin{equation}
\left\{ \psi _{i}^{1}\equiv \left\vert \phi _{i-}\right\rangle =\frac{1}{%
\sqrt{2}}(\left\vert 0\right\rangle -\left\vert i\right\rangle
),i=1,...,n-1\right\} ,  \label{state-}
\end{equation}%
where all $\left\vert i\right\rangle $ together with $\left\vert
0\right\rangle $\ form an orthonormal basis of the corresponding $n$%
-dimensional system $\beta $, then an interesting result occurs. That is,
the two corresponding measurements $M_{0}$ and $M_{1}$ can become
arbitrarily close to each other when $n$ increases.

As we know, the precision of any measurement device is restricted by the
quantum uncertainty principle. Consequently, when two measurements are
getting extremely close, they will eventually become indistinguishable.
Basing on these observations, in this paper we will build a QBC protocol
which can evade the no-go proofs.


In the next section, we will elaborate the conjunction between the HJW
theorem and the no-go proofs (i.e., the original MLC theorem and its
expansions) of unconditionally secure QBC. Then in section III, we will review
briefly the anomalous result found in \cite{HeSteering}. Our QBC protocol
will be presented in section IV, with its security against Alice and Bob
being proven in sections V and VI, respectively. Section VII gives a technical
remark on the mathematical method used in the security proof. The
significance of the result, especially its impact on the possible
development of fundamental theories, will be addressed in section VIII.

\section{No-go proofs and the HJW theorem}

Let us first review briefly the general features of the no-go proofs of QBC
\cite%
{qi74,qi24,qi23,qi56,qi58,qi105,qi82,qbc28,qi47,qi611,qi157,qi147,qbc49,qi101,qbc3,qbc36,qbc35,qi610,qi240,qi283,qi323,qi715,qbc12,qi714,qbc32,qbc31,qbc121,qbc140,qbc146,qbc148,qbc162}%
.

(I) The reduced model. Following the presentation of \cite{qi23}, in these
proofs any QBC protocol can be rephrased as the following general procedure.

\qquad (1) Alice prepares the state%
\begin{equation}
\left\vert \Omega _{b}\right\rangle =\sum\limits_{i}\lambda
_{i}^{b}\left\vert \zeta _{i}^{b}\right\rangle _{\alpha }\otimes \left\vert
\psi _{i}^{b}\right\rangle _{\beta }\text{ .}  \label{coding}
\end{equation}%
according to the value of $b$ that she wants to commit to Bob. Here, for
each specific value of $b$, $\left\vert \zeta _{i}^{b}\right\rangle _{\alpha
}$'s are a set of orthogonal states of system $\alpha $, while $\left\vert
\psi _{i}^{b}\right\rangle _{\beta }$'s are not necessarily orthogonal to
each other.

\qquad (2) An honest Alice is now supposed to make a measurement on $\alpha $
and determine the value of $i$.

\qquad (3) Alice sends system $\beta $ to Bob as the evidence for her
commitment. This completes the commit phase.

\qquad (4) In the unveil phase, Alice opens the commitment by declaring the
values of $b$ and $i$, i.e., she reveals which $\left\vert \psi
_{i}^{b}\right\rangle _{\beta }$\ was sent to Bob.

\qquad (5) Bob measures system $\beta $ to verify Alice's declared data.

(II)\ The concealing condition. To ensure that Bob's information on the
committed bit is trivial before the unveil phase, any QBC protocol secure
against Bob should satisfy%
\begin{equation}
\rho _{0}^{\beta }\simeq \rho _{1}^{\beta },  \label{concealing}
\end{equation}%
where $\rho _{b}^{\beta }\equiv Tr_{\alpha }\left\vert \Omega
_{b}\right\rangle \left\langle \Omega _{b}\right\vert $ is the reduced
density matrix of the state sent to Bob corresponding to Alice's committed
bit $b$.

(III) The cheating strategy. Here is where the HJW theorem comes to serve. A
dishonest Alice can begin the QBC protocol described by the reduced model by
preparing system $\alpha \otimes \beta $\ in such a state that $\beta $\
alone has density matrix $\rho _{0}^{\beta }$. Then she skips the
measurement in step (2) so that $\alpha $ and $\beta $\ remain entangled
throughout the commit phase. In the unveil phase, since Eq. (\ref{concealing}%
) is satisfied, the HJW theorem ensures that she can unveil the state of $%
\beta $ as either $\left\vert \psi _{i}^{0}\right\rangle _{\beta }$ or $%
\left\vert \psi _{i}^{1}\right\rangle _{\beta }$\ at her will, simply by
choosing between the two measurements $M_{0}$, $M_{1}$\ mentioned in the HJW
theorem (as summarized in Introduction) on $\alpha $.

In other words (as in the presentation in \cite{qi23}), Alice can start the
protocol with $\left\vert \Omega _{0}\right\rangle $ as described in Eq. (%
\ref{coding}) and skips step (2). Later if she wants to unveil $b=0$, she
simply measures $\alpha $ in the basis $\{\left\vert \zeta
_{i}^{0}\right\rangle _{\alpha }\}$ (equivalent to applying measurement $%
M_{0}$) in step (4). Or if she wants to unveil $b=1$, all she needs is to
apply a unitary transformation $U$ to rotate her measurement basis from $%
\{\left\vert \zeta _{i}^{0}\right\rangle _{\alpha }\}$ to $\{\left\vert
\zeta _{i}^{1}\right\rangle _{\alpha }\}$, or equivalently, apply the
transformation $U^{\dagger }$ on $\alpha $ which maps $\left\vert \Omega
_{0}\right\rangle $ into $\left\vert \Omega _{1}\right\rangle $, then still
measure it in the basis $\{\left\vert \zeta _{i}^{0}\right\rangle _{\alpha
}\}$ (both are equivalent to applying measurement $M_{1}$).

Consequently, Alice needs not to determine the value of $b$ until the unveil
phase. That is, a concealing QBC protocol cannot be binding, so that
unconditionally secure QBC deems impossible.

Note that the HJW theorem applies only to the case where $\rho _{0}^{\beta
}=\rho _{1}^{\beta }$ is satisfied rigorously. But the no-go proofs also
apply to protocols that are $\varepsilon $-concealing, i.e., satisfying $%
\rho _{0}^{\beta }\simeq \rho _{1}^{\beta }$\ instead of $\rho _{0}^{\beta
}=\rho _{1}^{\beta }$. The detailed cheating strategy is: Alice commits to
one of the density matrix (e.g., $\rho _{0}^{\beta }$) first. Then if she
wants to alter the commitment to $b=1$, she turns $\rho _{0}^{\beta }$ into
another density matrix $\rho _{1}^{\prime \beta }$ using the method
described in HJW theorem, where $\rho _{1}^{\prime \beta }=\rho _{0}^{\beta
} $, and the states corresponding to $\rho _{1}^{\prime \beta }$\ are very
close to those corresponding to $\rho _{1}^{\beta }$. Since $\rho
_{1}^{\prime \beta }\simeq \rho _{1}^{\beta }$, Bob can hardly distinguish
them apart, so that the cheating can be successful with a probability $%
\varepsilon $-close to $1$. That is, the cheating on $\varepsilon $%
-concealing protocols still relies on the HJW theorem. Dishonest Alice still
needs to find two \textit{different}\ measurements $M_{0}$ and $M_{1}$\ on $%
\alpha $ to steer Bob's system $\beta $ from $\rho _{0}^{\beta }$\ to $\rho
_{1}^{\prime \beta }$.

After the early appearance of the MLC no-go theorem \cite{qi74,qi24,qi23},
there were many newer no-go proofs \cite%
{qi56,qi58,qi105,qi82,qbc28,qi47,qi611,qi157,qi147,qbc49,qi101,qbc3,qbc36,qbc35,qi610,qi240,qi283,qi323,qi715,qbc12,qi714,qbc32,qbc31,qbc121,qbc140,qbc146,qbc148,qbc162}
which enriched the above reduced model in different aspects so that they can
be more general and rigorous. But the HJW theorem is always the base of the
final step of their cheating strategies, though sometimes not explicitly
cited.

In literature, there were attempts on building QBC which challenge either
the above feature (I) or (II) (see \cite{HeJPA,qbc75,HeQIP,qbc157} and the
references therein). But we will present a QBC protocol which satisfies the
features (I) and (II), while the cheating strategy (III) does not work. This
is based on a recent discovery on quantum steering, as reviewed below.

\section{Chaos in steering in high-dimensional systems}

Very recently, it was found that in high-dimensional systems, there exists a
specific form of bipartite quantum system which can display a kind of chaos
effect when being adopted for steering \cite{HeSteering}. That is, \textit{a
subtle difference in the measurement results on one side can steer the other
side into completely orthogonal states}.

More specifically, let us take $\psi _{i}^{0}$'s\ and $\psi _{i}^{1}$'s
mentioned in the above description of the HJW theorem as the two sets of
evenly distributed states
$\{\phi _{i+}\}$ and $\{\phi _{i-}\}$ in Eqs. (\ref{state+}) and (\ref{state-}),
and denote their density matrics as $\rho _{+}$\ and $\rho _{-}$,
respectively. It was proven in
\cite{HeSteering} that the trace distance between $\rho _{+}$
and $\rho _{-}$ is%
\begin{equation}
D(\rho _{+},\rho _{-})\equiv \frac{1}{2}tr\sqrt{(\rho _{+}-\rho _{-})^{\dag
}(\rho _{+}-\rho _{-})}=\frac{1}{\sqrt{n-1}}.  \label{trace distance}
\end{equation}%
Therefore, $\rho _{+}$\ and $\rho _{-}$\ can be arbitrarily close to each
other with the increase of $n$. But for any finite $n$, $\rho _{+}=\rho _{-}$
cannot be satisfied rigorously so that the HJW theorem cannot be applied
directly. Thus, Alice cannot expect to find a bipartite system $\alpha
\otimes \beta $\ such that her local measurements on $\alpha $ alone can
steer the state of $\beta $ from an element of $\left\{ \left\vert \phi
_{i+}\right\rangle \right\} $\ to an element of $\left\{ \left\vert \phi
_{i-}\right\rangle \right\} $ with a probability equals exactly to $1$. Now
let us study what happens if Alice tries to steer the state of $\beta $ to
another state which is very close to an element of $\left\{ \left\vert \phi
_{i-}\right\rangle \right\} $, as described above in Alice's cheating
strategy on $\varepsilon $-concealing QBC protocols.

Suppose that Alice prepares a bipartite system $\alpha \otimes \beta $ in
the state%
\begin{equation}
\left\vert \Omega \right\rangle =\frac{1}{\sqrt{n-1}}\sum_{i=1}^{n-1}\left%
\vert \alpha _{i+}\right\rangle _{\alpha }\left\vert \phi _{i+}\right\rangle
_{\beta }.  \label{omega+}
\end{equation}%
Here $\{\left\vert \alpha _{i+}\right\rangle _{\alpha },i=0,...,n-1\}$ is an
orthonormal basis of the $n$-dimensional system $\alpha $ (the subscripts $%
\alpha $ and $\beta $ will be omitted thereafter). Obviously there is:

\textit{Result 1: for any }$i$\textit{, if Alice projects }$\alpha $\textit{%
\ into }$\left\vert \alpha _{i+}\right\rangle $\textit{\ then }$\beta $%
\textit{\ will collapse into }$\left\vert \phi _{i+}\right\rangle $\textit{.}

Now we will try to find the measurement on $\alpha $ which can make $\beta $
collapse to a state close to $\left\vert \phi _{i-}\right\rangle $. Defining%
\begin{equation}
\left\vert \phi _{n-}\right\rangle \equiv \frac{1}{\sqrt{n}}\left(
\left\vert 0\right\rangle +\sum\nolimits_{i=1}^{n-1}\left\vert
i\right\rangle \right) ,  \label{fai n}
\end{equation}%
\begin{equation}
\left\vert \tilde{\alpha}_{n-}\right\rangle \equiv \frac{1}{\sqrt{n-1}}%
\sum_{i=1}^{n-1}\left\vert \alpha _{i+}\right\rangle ,  \label{alfa n-}
\end{equation}%
and%
\begin{equation}
\left\vert \tilde{\alpha}_{i-}\right\rangle \equiv \frac{1}{\sqrt{1-4/n^{2}}}%
\left( \frac{2-n}{n}\left\vert \alpha _{i+}\right\rangle +\frac{2}{n}%
\sum_{i^{\prime }=1,i^{\prime }\neq i}^{n-1}\left\vert \alpha _{i^{\prime
}+}\right\rangle \right)  \label{alfa-}
\end{equation}%
for\ $i=1,...,n-1$, then Eq. (\ref{omega+}) equals to%
\begin{eqnarray}
\left\vert \Omega \right\rangle &=&\frac{1}{\sqrt{n-1}}\sum_{i=1}^{n-1}\sqrt{%
1-4/n^{2}}\left\vert \tilde{\alpha}_{i-}\right\rangle \left\vert \phi
_{i-}\right\rangle  \nonumber \\
&&+\sqrt{\frac{2}{n}}\left\vert \tilde{\alpha}_{n-}\right\rangle \left\vert
\phi _{n-}\right\rangle .  \label{omega-}
\end{eqnarray}

For a given $i\in \{1,...,n-1\}$, if Alice can project system $\alpha $ to $%
\left\vert \tilde{\alpha}_{i-}\right\rangle $, then Eq. (\ref{omega-}) shows
that system $\beta $ will collapse to%
\begin{eqnarray}
\left\vert \tilde{\phi}_{i-}\right\rangle &\equiv &c^{\prime }\left[ \frac{%
\left\vert \phi _{i-}\right\rangle +\sum_{i^{\prime }=1,i^{\prime }\neq
i}^{n-1}\left\langle \tilde{\alpha}_{i-}\right. \left\vert \tilde{\alpha}%
_{i^{\prime }-}\right\rangle \left\vert \phi _{i^{\prime }-}\right\rangle }{%
\sqrt{n-1}/\sqrt{1-4/n^{2}}}\right.  \nonumber \\
&&\left. +\sqrt{\frac{2}{n}}\left\langle \tilde{\alpha}_{i-}\right.
\left\vert \tilde{\alpha}_{n-}\right\rangle \left\vert \phi
_{n-}\right\rangle \right]  \nonumber \\
&=&c^{\prime }\left[ \frac{\sqrt{1-4/n^{2}}}{\sqrt{n-1}}\left( \left\vert
\phi _{i-}\right\rangle -\frac{4\sum_{i^{\prime }=1,i^{\prime }\neq
i}^{n-1}\left\vert \phi _{i^{\prime }-}\right\rangle }{n^{2}-4}\right)
\right.  \nonumber \\
&&\left. +\frac{\sqrt{2/n}\sqrt{n-2}}{\sqrt{n-1}\sqrt{n+2}}\left\vert \phi
_{n-}\right\rangle \right] ,  \label{beta final2}
\end{eqnarray}%
where%
\begin{equation}
c^{\prime }=\sqrt{\frac{n(n-1)(n+2)}{(n^{2}+2)}}.
\end{equation}%
Thus we obtain:

\textit{Result 2: for any }$i$\textit{, if Alice can project }$\alpha $%
\textit{\ into }$\left\vert \tilde{\alpha}_{i-}\right\rangle $\textit{\
defined in Eq. (\ref{alfa-}), then }$\beta $\textit{\ will collapse into }$%
\left\vert \tilde{\phi}_{i-}\right\rangle $\textit{\ in Eq. (\ref{beta
final2}).}

Multiplying $\left\langle \phi _{i-}\right\vert $\ by Eq. (\ref{beta final2}%
), we have
\begin{equation}
\left\langle \phi _{i-}\right. \left\vert \tilde{\phi}_{i-}\right\rangle =%
\sqrt{1-\frac{2n+2}{n^{2}+2}},  \label{new eq25}
\end{equation}%
i.e., $\left\vert \tilde{\phi}_{i-}\right\rangle $ is indeed very close to $%
\left\vert \phi _{i-}\right\rangle $.

Now let us study the relationship between the states in Results 1 and 2.
From Eq. (\ref{alfa-}) we find%
\begin{equation}
|\left\langle \alpha _{i+}\right. \left\vert \tilde{\alpha}%
_{i-}\right\rangle |^{2}=1-\frac{4}{n+2},  \label{old eq31}
\end{equation}%
i.e., $\left\vert \alpha _{i+}\right\rangle $ and $\left\vert \tilde{\alpha}%
_{i-}\right\rangle $ are very close to each other when $n$ is high. In
contrast, multiplying $\left\langle \phi _{i+}\right\vert $\ by the
right-hand side of Eq. (\ref{beta final2}), we have%
\begin{equation}
\left\langle \phi _{i+}\right. \left\vert \tilde{\phi}_{i-}\right\rangle =0
\end{equation}%
for any $n$, i.e., $\left\vert \phi _{i+}\right\rangle $ and $\left\vert
\tilde{\phi}_{i-}\right\rangle $ are always strictly orthogonal to each
other. Therefore, combining Results 1 and 2, we obtain the conclusion
mentioned at the beginning of this subsection, that for the bipartite state $%
\left\vert \Omega \right\rangle $ in Eq. (\ref{omega+}), a subtle difference
in the measurement results on $\alpha $ can steer $\beta $ into completely
orthogonal states.

Note that when $n$ is finite, for any $i\neq i^{\prime }$, Eq. (\ref{alfa-})
shows that $\left\langle \tilde{\alpha}_{i^{\prime }-}\right. \left\vert
\tilde{\alpha}_{i-}\right\rangle \neq 0$, so that $\{\left\vert \tilde{\alpha%
}_{i-}\right\rangle ,i=0,...,n-1\}$ cannot be used as an orthogonal
measurement basis. Thus it generally takes a positive-operator valued
measure (POVM) to project $\alpha $ to an element of $\{\left\vert \tilde{%
\alpha}_{i-}\right\rangle \}$. But with the fact that $\left\vert \alpha
_{i+}\right\rangle $ and $\left\vert \tilde{\alpha}_{i-}\right\rangle $ are
very close to each other, it is clear that the measurement/POVM $M_{+}$ and $%
M_{-}$\ for projecting $\alpha $ to an element of $\{\left\vert \alpha
_{i+}\right\rangle \}$ or $\{\left\vert \tilde{\alpha}_{i-}\right\rangle \}$%
, respectively, also become arbitrarily close to each other as $n$
increases. This is one of the key feature that leads to our QBC protocol.

\section{Our protocol}

Since proving that secure QBC exists even only in theory already has great
significance (as indicated by the Clifton-Bub-Halvorson (CBH) theorem that
we will addressed in the Discussion section), here for simplicity, we only
consider the ideal case without practical imperfections, such as
transmission errors, detection loss or dark counts, etc. Under this setting,
we propose the following protocol.

\bigskip

\textit{Our QBC protocol:}

\textit{The commit phase:}

\textit{(i) Alice decides on the value of }$b$\textit{\ that she wants to
commit. Then for }$j=1$\textit{\ to }$s$\textit{:}

\textit{She randomly picks }$i_{j}\in \{1,2,...,n_{B}\}$\textit{\ where }$%
n_{B}\rightarrow \infty $\textit{, and sends Bob a quantum register }$\Psi
_{j}$\textit{, which\ is an infinite-dimensional system prepared in the
state }$\psi _{i_{j}}^{b}=(\left\vert 0\right\rangle +(-1)^{b}\left\vert
i_{j}\right\rangle )/\sqrt{2}$\textit{.}

\textit{Note that in each round, }$i_{j}$\textit{\ is independently chosen,
while }$b$\textit{\ remains the same for all }$j$\textit{.}

\textit{(ii) Bob stores these }$s$\textit{\ quantum registers unmeasured.}

\textit{The unveil phase:}

\textit{(iii) Alice announces the values of }$b$\textit{\ and all }$i_{j}$%
\textit{\ (}$j=1,...,s$\textit{).}

\textit{(iv) Bob tries to project each }$\Psi _{j}$\textit{\ into the state }%
$\psi _{i_{j}}^{b}=(\left\vert 0\right\rangle +(-1)^{b}\left\vert
i_{j}\right\rangle )/\sqrt{2}$\textit{. If the projections are successful
for all registers, Bob accepts Alice's commitment. Else if any of the
projections fails, Bob concludes that Alice cheated.}

\bigskip

In brief, the protocol can be secure against Alice's cheating, as long as
the quantum uncertainty principle puts an upper bound on the precision
of her measurement devices, so that she cannot discriminate quantum states
with very little difference, which was required for implementing the attack
based on the HJW theorem against the specific states
$\{\psi _{i}^{0}\}$\ and $\{\psi _{i}^{1}\}$ in Eqs. (\ref{state+}) and (\ref{state-}). Also, the protocol is secure against dishonest
Bob, because the density matrices corresponding to $\{\psi _{i}^{0}\}$\ and $%
\{\psi _{i}^{1}\}$ become arbitrarily close to each other when the dimension
of the quantum system is sufficiently high. Now let us give the security proof in details.

\section{Security against dishonest Alice}

It is trivial to show that in the commit phase of our QBC protocol, if Alice
sends Bob each register $\Psi _{j}$ honestly in a pure state $\psi
_{i_{j}}^{b}=(\left\vert 0\right\rangle +(-1)^{b}\left\vert
i_{j}\right\rangle )/\sqrt{2}$ non-entangled with any other system, then she
cannot unveil the committed bit as $\bar{b}$. This is because $|(1/\sqrt{2}%
)(\left\langle 0\right\vert +(-1)^{\bar{b}}\left\langle i_{j}^{\prime
}\right\vert )(\left\vert 0\right\rangle +(-1)^{b}\left\vert
i_{j}\right\rangle )/\sqrt{2}|^{2}=\delta _{ii^{\prime }}/4$, so that no
matter how Alice chooses the value of $i_{j}^{\prime }$ of each $j$, the
total probability for her to announce the states of all the $s$ registers as
$\psi _{i_{j}^{\prime }}^{\bar{b}}$ instead of the actual $\psi _{i_{j}}^{b}$
without being caught is bounded by $(1/4)^{s}$, which is arbitrarily close
to $0$ for a sufficiently high $s$ value.

Now consider Alice's general attack using entangled states. For each
register $\Psi _{j}$ sent to Bob, the state of Alice's and Bob's combined
system $\alpha \otimes \beta $ can always be written as%
\begin{equation}
\left\vert \Omega _{j}\right\rangle =\sum_{i_{j}=1}^{n_{A}-1}\lambda
_{i_{j}}\left\vert \alpha _{i_{j}+}\right\rangle _{\alpha }\left\vert \beta
_{i_{j}+}\right\rangle _{\beta },
\end{equation}%
where $\lambda _{i_{j}}$ denotes the superposition coefficient. We must
emphasize that this form covers all possible states that dishonest Alice may
use. For example, even if she entangles different registers $\Psi _{1}$, $%
\Psi _{2}$, ..., $\Psi _{j}$, ... together, we can still single out $\Psi
_{j}$ as Bob's system $\beta $ in this equation, while treating all other $%
\Psi _{j^{\prime }}$ ($j^{\prime }\neq j$) as a part of Alice's system $%
\alpha $. We also assume that she has full control over system $\alpha $
(although in fact she could not do so if $\alpha $\ includes other $\Psi
_{j^{\prime }}$), so that the security analysis below covers the upper bound
of Alice's cheating probability.

In this case, if $\{\left\vert \beta _{i_{j}+}\right\rangle _{\beta
},i_{j}=1,...,n_{A}-1\}$ contains any element not belonging to $\left\{ \psi
_{i}^{0}=\left\vert \phi _{i+}\right\rangle ,i=1,...,n-1\right\} $\ defined
in Eq. (\ref{state+}), and/or $\{\left\vert \alpha _{i_{j}+}\right\rangle
_{\alpha },i_{j}=1,...,n_{A}-1\}$ is not an orthogonal basis of $\alpha $,
then Alice cannot always unveil the state of $\Psi _{j}$ as $\psi _{i}^{0}$
with the correct $i$ value. Suppose that the error rate is $\eta $, then for
all the $s$ registers, her probability of unveiling $b=0$ successfully is at
the order of magnitude of $(1-\eta )^{s}$, which is trivial for high $s$.
Therefore, to ensure that she can unveiling $b=0$ without being caught, $%
\{\left\vert \alpha _{i_{j}+}\right\rangle _{\alpha }\}$ has to be chosen as
an orthogonal basis of $\alpha $, and $\{\left\vert \beta
_{i_{j}+}\right\rangle _{\beta }\}$ must be $n_{A}-1$ elements selected from
$\left\{ \left\vert \phi _{i+}\right\rangle ,i=1,...,n-1\right\} $. Without
loss of generality, here we suppose that she chooses the first $n_{A}-1$
ones in order, i.e., $\{\left\vert \beta _{i_{j}+}\right\rangle _{\beta
}=\left\vert \phi _{i_{j}+}\right\rangle ,i_{j}=1,...,n_{A}-1\}$, and each $%
\left\vert \phi _{i_{j}+}\right\rangle $ is chosen with the equal
probability. Omitting the subscript\ $j$, we have%
\begin{equation}
\left\vert \Omega \right\rangle =\frac{1}{\sqrt{n_{A}-1}}%
\sum_{i=1}^{n_{A}-1}\left\vert \alpha _{i+}\right\rangle _{\alpha
}\left\vert \phi _{i+}\right\rangle _{\beta }.
\end{equation}

This equation is much the same as Eq. (\ref{omega+}), except that $n$ is
replaced by $n_{A}$. Thus the analysis in the previous subsection still
applies. That is, although this $\left\vert \Omega \right\rangle $ can
ensure Alice to unveil $b=0$ with probability $100\%$, it cannot be unveiled
as $b=1$ without error. This is because Eq. (\ref{omega-}) shows that system
$\beta $ has a probability $2/n_{A}$ to be projected into $\left\vert \phi
_{n_{A}-}\right\rangle =(\left\vert 0\right\rangle
+\sum\nolimits_{i=1}^{n_{A}-1}\left\vert i\right\rangle )/\sqrt{n_{A}}$,
which is not a legitimate state for committing $b=1$, and it is also
orthogonal to all legitimate states $\left\vert \phi _{i-}\right\rangle $\ ($%
i=1,...,n_{A}-1$). Then Alice's announcing $b=1$ stands at least the
probability $2/n_{A}$ to be caught cheating for each register, and the total
probability for her to pass Bob's check on all the $s$ registers is not
greater than%
\begin{equation}
p\equiv (1-2/n_{A})^{s},  \label{p}
\end{equation}%
which drops exponentially to $0$ as $s$ increases.

Also, Eq. (\ref{p}) is merely a loose upper bound because as we mentioned, $%
\{\left\vert \tilde{\alpha}_{i-}\right\rangle \}$ in Eq. (\ref{omega-}) is
not an orthogonal basis for any finite $n$. Meanwhile, Eq. (\ref{new eq25})
shows that $\left\vert \tilde{\phi}_{i-}\right\rangle $ does not equal to $%
\left\vert \phi _{i-}\right\rangle $ exactly when $n$ is finite. Therefore,
it is impossible for Alice to discriminate unambiguously which $\left\vert
\phi _{i-}\right\rangle $ ($i=1,...,n$) is the one that Bob's system $\beta $
will collapse to, so that Alice's actual probability for passing Bob's check
on all the $s$ registers will be even smaller than Eq. (\ref{p}).

More importantly, due to the existence of quantum uncertainty principle, any
measurement device cannot be adjusted with unlimited precision, so that it
cannot have unlimited power on discriminating quantum states that are very
close to each other. If Alice chooses an$\ $extremely high\ $n_{A}$ value, $%
\left\vert \alpha _{i+}\right\rangle $ and $\left\vert \tilde{\alpha}%
_{i-}\right\rangle $ can become so close (as shown by Eq. (\ref{old eq31})),
that no physical device in the world can help Alice distinguish them apart.
Consequently, whether $\beta $ collapses to $\left\vert \phi
_{i+}\right\rangle $ or $\left\vert \phi _{i-}\right\rangle $ becomes
completely out of her control. That is, for any physical implementation of
our protocol, the uncertainty principle puts a limit on the maximum of the $%
n_{A}$ value that Alice can choose (denoted as $n_{A\max }$), such that she
can discriminate $\left\vert \alpha _{i+}\right\rangle $ and $\left\vert
\tilde{\alpha}_{i-}\right\rangle $ and thus steer Bob's system $\beta $ only
if she chooses $n_{A}\leq n_{A\max }$. But she cannot do so anymore if she
chooses $n_{A}>n_{A\max }$. As a result, basing on the closest $\left\vert
\alpha _{i+}\right\rangle $ and $\left\vert \tilde{\alpha}_{i-}\right\rangle
$ that can be discriminated by the measurement devices actually used, we can
determine $n_{A\max }$ beforehand according to Eq. (\ref{old eq31}) (the
exact value will depend on the specific implementation scheme though, so we
cannot have a general estimation here). Then with Eq. (\ref{p}), we know
that for any expected value $p_{A\max }$, by choosing%
\begin{equation}
s\geq \frac{\ln p_{A\max }}{\ln (1-2/n_{A\max })},
\end{equation}%
it is sufficient to guarantee that the probability of Alice's successful
cheating is bounded by $p\leq p_{A\max }$.

\section{Security against dishonest Bob}

As mentioned above, it was proven in Ref. \cite{HeSteering} that the density
matrices $\rho _{+}$, $\rho _{-}$ for $\{\psi _{i}^{0}\}$, $\{\psi _{i}^{1}\}
$ defined in Eqs. (\ref{state+}) and (\ref{state-}), respectively, satisfy
Eq. (\ref{trace distance}). Also, the $n_{B}$ value in our protocol is
irrelevant with dishonest Alice's $n_{A}$ and can be chosen to be much
higher than $s$. Therefore, from Bob's point of view the range of Alice's
selected $i_{j}$ for each $\psi _{i_{j}}^{b}$ is always $\{1,...,n_{B}\}$
with $n_{B}\rightarrow \infty $ so that $\rho _{+}$ and $\rho _{-}$ are
arbitrarily close to each other. The commit phase of our above protocol is
simply to send such states $s$ times. Thus the density matrices of the
states sent to Bob for committing $b=0$ and $b=1$, respectively, are $\rho
_{0}^{\beta }=\rho _{+}^{\otimes s}$ and $\rho _{1}^{\beta }=\rho
_{-}^{\otimes s}$. Obviously, the concealing condition
Eq. (\ref{concealing}) is satisfied. Therefore, the states are completely
indistinguishable to Bob before the unveil phase, so that the protocol is
perfectly secure against his cheating.

\section{A technical remark on taking the $n\rightarrow \infty $\ limit}

Some might wonder whether it is legitimate to take the $n\rightarrow \infty $%
\ limit in Eq. (\ref{trace distance}) and the related equations in Ref. \cite%
{HeSteering}, because we are studying the two sets of evenly distributed
states defined in Eqs. (\ref{state+}) and (\ref{state-}), where each state
occurs with the equal probability $1/(n-1)$. When $n\rightarrow \infty $\
this probability becomes $0$ which does not seem to make sense. But we
should always keep in mind that the actual question related with our QBC
protocol is: when Alice randomly selects a state from $\left\{ \left\vert
\phi _{i+}\right\rangle ,i=1,...,n-1\right\} $\ or $\left\{ \left\vert \phi
_{i-}\right\rangle ,i=1,...,n-1\right\} $, can Bob distinguish which
set it is from? Surely Alice can still make such a selection when $%
n\rightarrow \infty $, and each state can be picked with the same
probability even though we may not write this probability as $1/(n-1)$.
Also, the fact that the trace distance satisfies $D(\rho _{+},\rho _{-})=1/%
\sqrt{n-1}$ for any finite $n$ clearly shows that distinguishing $\rho _{+}$
and $\rho _{-}$ becomes harder and harder as $n$ increases. Thus it is
natural to conclude that it will be even harder to distinguish whether a
state is picked from $\left\{ \left\vert \phi _{i+}\right\rangle \right\} $\
or $\left\{ \left\vert \phi _{i-}\right\rangle \right\} $ when $n\rightarrow
\infty $. Therefore the security of our protocol against dishonest Bob stays
valid no matter the above density matrix description of the states is
adopted or not.

\section{Discussion}

Although we merely show the theoretical existence of unconditionally secure
QBC without studying its feasibility in practice, the result is still very
important. It re-opens the venues for cryptographic tasks that once closed
by the no-go proofs based on the HJW theorem, such as quantum coin flipping
and two-party secure computations \cite{qi149}.

But more importantly, it also contributes to the development of fundamental
theories. There is an interesting result called the CBH theorem \cite{qi256}%
, which is an inspiring attempt to raise some information-theoretic
constraints to the level of fundamental laws of Nature, from which quantum
theory can be deduced. These constraints were suggested to be three
\textquotedblleft no-go's\textquotedblright , which are (I) the
impossibility of superluminal information transfer, (II) the impossibility
of perfectly broadcasting of an unknown state, and (III) the impossibility
of unconditionally secure BC. But recently, Heunen and Kissinger \cite%
{qbc135} suggested that the impossibility of BC is not caused by the
conceptual structure of quantum theory, but by the algebraic model assumed
in \cite{qi256}. Logically, this may indicates that the fundamental axioms of
quantum theory \textit{alone} do not necessarily lead to impossibility of
BC. Thus our finding (that QBC can be secure in infinite-dimensional
systems) seems to be in good agreement with this result. Therefore, we may need to seek for another information-theoretic principle as the third constraint in
the CBH theorem. Or we will have to add \textquotedblleft strictly
infinite-dimensional systems do not exist\textquotedblright\ as an
additional axiom to keep the no-go proofs of QBC valid.

Manipulating infinite-dimensional systems may indeed be hard in practice if
we want to use physical systems with an infinite number of energy levels,
because it may imply an infinitely high energy. But there could be tricks to
use other degree of freedoms to serve as a replacement. We will study such
practical implementations in successive works.



\end{document}